\documentclass[10pt,a4paper,twoside]{article}
\usepackage{epsfig}
\usepackage{baltlat6}
\usepackage{array}
\usepackage{here}
\begin{document}
\ \
\vspace{0.5mm}
\setcounter{page}{277}

\titlehead{Baltic Astronomy, vol.\,24, 000--000, 2015}

\titleb{ASTRONOMICAL SURVEYS AND BIG DATA}

\begin{authorl}
\authorb{Areg M. Mickaelian}{1}
\end{authorl}

\begin{addressl}
\addressb{1}{NAS RA V. Ambartsumian Byurakan Astrophysical Observatory (BAO), \\ Byurakan 0213, Aragatzotn Province, Armenia; aregmick@yahoo.com}
\end{addressl}

\submitb{Received: 2015 September 1; accepted: 2015 September}

\begin{summary} Recent all-sky and large-area astronomical surveys and their catalogued data over the whole range of electromagnetic spectrum are reviewed, from $\gamma$-ray to radio, such as Fermi-GLAST and INTEGRAL in $\gamma$-ray, ROSAT, XMM and Chandra in X-ray, GALEX in UV, SDSS and several POSS I and II based catalogues (APM, MAPS, USNO, GSC) in optical range, 2MASS in NIR, WISE and AKARI IRC in MIR, IRAS and AKARI FIS in FIR, NVSS and FIRST in radio and many others, as well as most important surveys giving optical images (DSS I and II, SDSS, etc.), proper motions (Tycho, USNO, Gaia), variability (GCVS, NSVS, ASAS, Catalina, Pan-STARRS) and spectroscopic data (FBS, SBS, Case, HQS, HES, SDSS, CALIFA, GAMA). An overall understanding of the coverage along the whole wavelength range and comparisons between various surveys are given: galaxy redshift surveys, QSO/AGN, radio, Galactic structure, and Dark Energy surveys. Astronomy has entered the Big Data era. Astrophysical Virtual Observatories and Computational Astrophysics play an important role in using and analysis of big data for new discoveries.
\end{summary}


\resthead{Astronomical Surveys and Big Data}
{A. M. Mickaelian}

\sectionb{1}{INTRODUCTION}

Astronomical surveys are the main source for discovery of astronomical objects and accumulation of observational data for further analysis, interpretation, and achieving scientific results. Recent large multiwavelength (MW) surveys both by ground-based and space telescopes and their catalogues during the last 15 years have accumulated vast amounts of data over the whole range of electromagnetic spectrum from $\gamma$-ray to radio. Present astronomical databases and archives contain billions of objects, both Galactic and extragalactic, and the vast amount of data on them allow new studies and discoveries. The Big Data era has come. Astrophysical Virtual Observatories (VO) use available databases and current observing material as a collection of interoperating data archives and software tools to form a research environment in which complex research programs can be conducted. Most of the modern databases give at present VO access to the stored information. This makes possible not only the open access but also a fast analysis and managing of these data. VO is a prototype of Grid technologies that allows distributed data computation, analysis and imaging. Particularly important are data reduction and analysis systems: spectral analysis, spectral energy distribution (SED) building and fitting, modelling, variability studies, cross-correlations, etc. Numerical or Computational Astrophysics (part of Computer Science, also called Laboratory Astrophysics) has become an indissoluble part of astronomy and most of modern research is being done by means of it. Very often dozens of thousands of sources hide a few very interesting ones that are needed to be discovered by comparison of various physical characteristics. Cross-correlations result in revealing new objects and new samples. The large amount of data requires new approaches to data reduction, management and analysis. Powerful computer technologies are required, including clusters and grids. Large volume astronomical servers have been established to host Big Data and giving high importance to their maintenance, the International Council of Scientific Unions (ICSU) has recently created World Data System (WDS) to unify data coming from different science fields for further possibility of exchange and new science projects.

In our review paper (Mickaelian 2012) ``Large Astronomical Surveys, Catalogs and Databases" we summarized astronomical data accumulated during the recent decades and presented astronomical surveys, catalogues, archives, databases and VOs. During these 3 years, many new ground-based and space surveys provided new data and many new releases appeared. Therefore, here we give an update of this information especially emphasizing the importance and going into details of astronomical surveys, as MW photometric ones from gamma-ray to radio, so as proper motion, variability and spectroscopic surveys, including objective prism low-dispersion surveys and digital ones.

\sectionb{2}{MULTIWAVELENGTH SURVEYS AND CATALOGUES}

Multiwavelength studies significantly changed our views on cosmic bodies and phenomena, giving an overall understanding and posiibility to combine and/or compare data coming from various wavelength ranges. MW astronomy appeared during the last few decades and recent MW surveys (including those obtained with space telescopes) led to catalogues containing billions of objects along the whole electromagnetic spectrum. When combining MW data, one can learn much more due to variety of information related to the same object or area, such as e. g. the Milky Way (Fig 1).


\begin{figure}[!tH]
\vbox{
\centerline{\psfig{figure=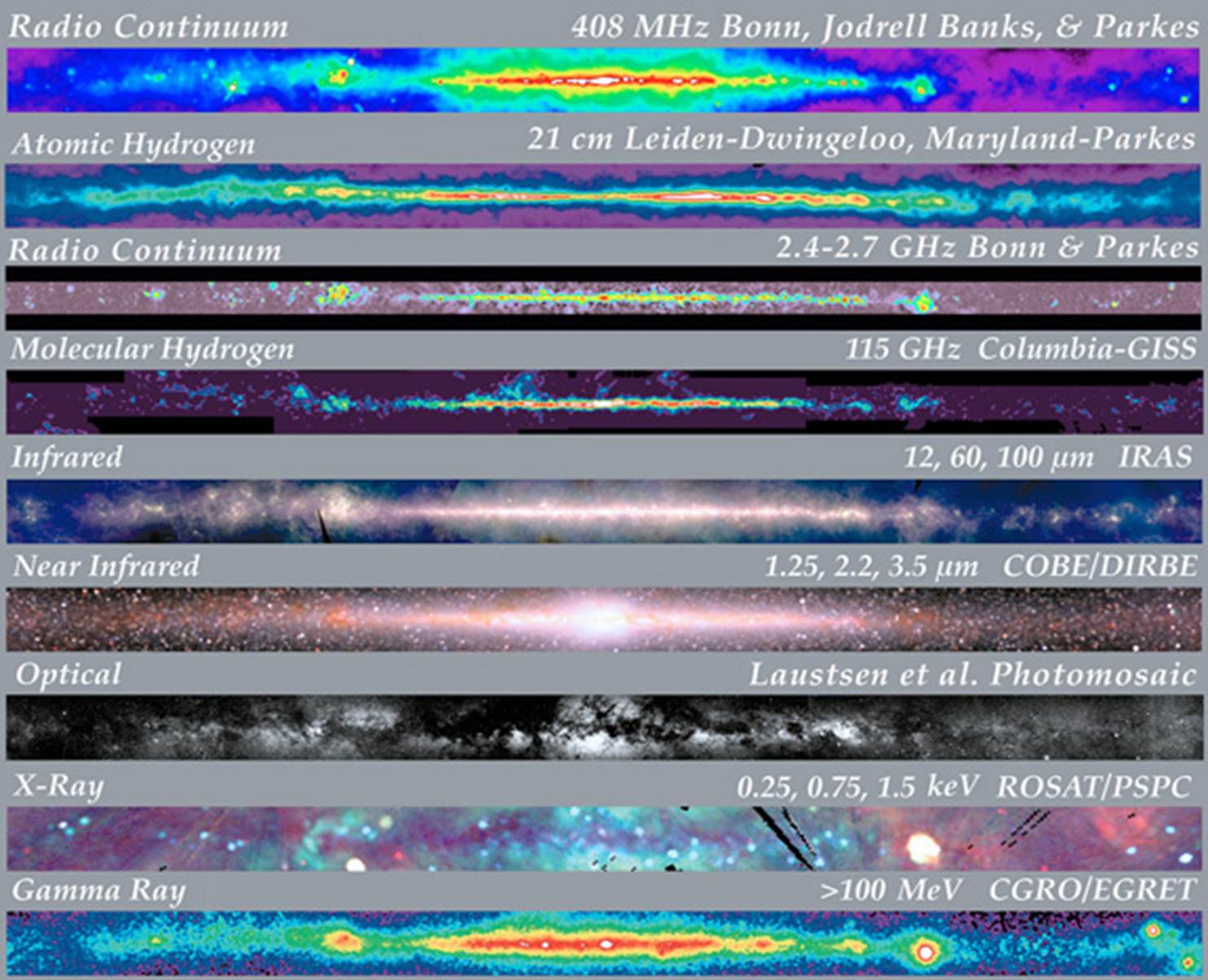,width=70mm,angle=0,clip=}}
\vspace{1mm}
\captionb{1}
{Milky Way image obtained in various wavelengths from $\gamma$-ray to radio and showing differences both in individual objects/formations and in the overall shape.}
}
\end{figure}

Here we list most important (having homogeneous data for a large number of sources over large area) recent surveys and resulted catalogues providing photometric data along the whole wavelength range:  

\begin{itemize}

\item \emph{\textbf{$\gamma$-ray.}} \textbf{Fermi-GLAST} 3FGL catalogue: gamma-ray positions and 10 MeV – 100 GeV photon counts for 3,033 sources (Acero et al. 2015); \textbf{INTEGRAL}: IBIS/ISGRI soft gamma-ray survey catalog of 1,126 sources (Bird et al. 2010); older \textbf{CGRO EGRET}: gamma-ray positions and 20 keV – 30 GeV photon counts for 1300 sources, including only 271 identified ones (Hartman et al. 1999); \textbf{Swift}: survey in deep fields, 9387 sources and \textbf{BeppoSAX}: 1082 gamma-ray bursts and some other data are much more accurate, however no all-sky or large-area catalogue is available from these missions/telescopes; typically gamma-ray sources are difficult to identify due to inaccurate positions (several arcmin errors); 

\item \emph{\textbf{X-ray.}} \textbf{R\"ontgensatellit (ROSAT) BSC}: X-ray positions and 0.07–2.4 keV photon counts and two hardness ratios for 18,806 sources (Voges et al. 1999); \textbf{ROSAT FSC}: X-ray positions and 0.07–-2.4 keV photon counts and two hardness ratios for 105,924 sources (Voges et al. 2000), ROSAT sources are difficult to identify due to inaccurate positions ($\sim$1$\arcmin$ errors); \textbf{INTEGRAL}: hard X-ray all-sky survey catalog of 403 sources; older \textbf{EXOSAT} Medium energy (1–8 keV) Slew Survey Catalog (EXMS): 1210 sources, including 992 identified ones; \textbf{ASCA}: 1190 sources and other data are available; much more accurate recent \textbf{X-ray Multi-Mirror mission (XMM-Newton)}: 372,728 sources from various surveys and \textbf{Chandra}: 380,000 sources from various surveys (no all-sky or large-area coverage); 

\item \emph{\textbf{Ultraviolet (UV).}} \textbf{Galaxy Evolution Explorer (GALEX) AIS}: 65,266,291 sources and \textbf{GALEX MIS}: 12,597,912 sources, 2$\arcsec$ rms accurate positions and fluxes in FUV 1528\AA ~and NUV 2271\AA ~bands for some 78 million sources over the whole sky up to 20.8$^{m}$ for AIS and 22.7$^{m}$ for MIS (Bianchi et al. 2011); a number of older mission data are available; \textbf{EUVE}: 514 sources; \textbf{FUSE}: FUV spectra of $\sim$3000 objects and others with much less accuracy and sensitivity; \textbf{Hubble Space Telescope (HST)}: 2941\AA ~NUV observations (no large area survey is available); 

\item \emph{\textbf{Optical range.}} \textbf{Sloan Digital Sky Survey (SDSS)} Data Release 12 (DR12): 67 mas rms positions, u, g, r, i, z photometry for 932,891,133 objects mainly in Northern extragalactic sky, limited to 22.0$^{m}$ in u, 22.2$^{m}$ in g and r, 21.3$^{m}$ in i and 20.5$^{m}$ in z (Alam et al. 2015); several POSS I and POSS II based catalogues; \textbf{USNO-A2.0}: 526,280,881 objects (Monet et al. 1998), still useful as photometric data differ from USNO-B1.0 and sometimes are better; \textbf{Cambridge Automated Plate Measurement (APM)}: 0.6$\arcsec$ rms positions, b and r photometry for 166,466,987 POSS I objects over 21,000 deg$^{2}$, limited to 21.0$^{m}$ in b and 20.0$^{m}$ in r (McMahon et al. 2000); \textbf{Minnesota Automated Plate Scanner (MAPS)}: 0.5$\arcsec$ rms positions, O/E photometry for 89,234,404 POSS-I objects over 21,000 deg$^{2}$, limited to 21.0$^{m}$ in O and 20.0$^{m}$ in E (Cabanela et al. 2003); \textbf{USNO-B1.0}: 0.4$\arcsec$ rms positions, B1, R1, B2, R2, and I photometry; proper motions for all 1,045,913,669 objects present in POSS-I/SERC-J and POSS-II/AAO-SES over the whole sky, limited to 22.5$^{m}$ in B and 20.8$^{m}$ in R; so far the largest among all available catalogues by the number of objects (Monet et al. 2003); and \textbf{GSC 2.3.2}: 0.3$\arcsec$ rms positions, j/V/F/N photometry for 945,592,683 objects present in DSS2, limited to 22.5$^{m}$ in j and 20.8$^{m}$ in F (Lasker et al. 2008); \textbf{Tycho-2}: 85 mas rms positions and accurate B/V magnitudes for 2,539,913 objects over the whole sky, limited to 15.2$^{m}$ in B and 16.3$^{m}$ in V (Hog et al. 2000);

\item \emph{\textbf{Near-Infrared (NIR).}} \textbf{2MASS Point Source Catalog (PSC)}: 0.5$\arcsec$ rms accurate positions and JHKs photometry for 470,992,970 objects over the whole sky up to 17.1$^{m}$ in J, 16.4$^{m}$ in H and 15.3$^{m}$ in Ks (Cutri et al. 2003); \textbf{2MASS Extended Source Catalog (ESC)}: JHK photometry for 1,647,599 objects over the whole sky (Skrutskie et al. 2006); \textbf{DEep Near Infrared Survey (DENIS)}: IJKs photometry for 355,220,325 objects over 16,700 deg$^{2}$ of the Southern sky up to 18.5$^{m}$ in I, 16.5$^{m}$ in J and 14.0$^{m}$ in Ks (DENIS 2005); deeper NIR observations (no large-area survey) from VLT, UKIRT, and other large IR telescopes;

\item \emph{\textbf{Mid-Infrared (MIR).}} \textbf{Wide-field Infrared Survey Explorer (WISE)}: 0.5$\arcsec$ rms accurate positions and 3.4 $\mu$m (80 $\mu$Jy sensitivity), 4.6 $\mu$m (110 $\mu$Jy), 12 $\mu$m (1 mJy), 22 $\mu$m (6 mJy) bands photometry for 563,921,584 sources over the whole sky (Cutri et al. 2012); \textbf{AKARI-IRC} Point Source Catalogue: 0.3$\arcsec$ rms accurate positions and 9 $\mu$m (50 mJy sensitivity) and 18 $\mu$m (120 mJy sensitivity) fluxes for 870,973 sources over the whole sky (Ishihara et al. 2010); \textbf{InfraRed Astronomical Satellite Point Source Catalog (IRAS PSC)} and \textbf{Faint Source Catalog (IRAS FSC)}: IR positions and fluxes in 12 $\mu$m (0.4 Jy sensitivity) and 25 $\mu$m (0.5 Jy) bands for 245,889 sources (IRAS 1988) and 173,044 sources (Moshir et al. 1990), respectively; as well as IRAS joint catalogue has been compiled containing 345,162 sources (Abrahamyan et al. 2015); \textbf{IRAS SSSC}: IR positions and fluxes in 12 $\mu$m and 25 $\mu$m bands for 16,740 extended sources over the whole sky (Helou et al. 1985); \textbf{Spitzer Space Telescope (SST) }7-band data (altogether 4,261,028 sources observed) have accurate positions and much higher sensitivity, however no all-sky or large-area survey is available from this mission; 

\item \emph{\textbf{Far-infrared (FIR).}} \textbf{IRAS PSC} and \textbf{IRAS FSC}: IR positions and fluxes in 60 $\mu$m (0.6 Jy sensitivity) and 100 $\mu$m (1.0 Jy) bands for 245,889 sources (IRAS 1988) and 173,044 sources (Moshir et al. 1990), respectively; 345,162 sources in total (Abrahamyan et al. 2015); \textbf{IRAS SSSC}: IR positions and fluxes in 60 $\mu$m and 100 $\mu$m bands for 16,740 extended sources over the whole sky (Helou et al. 1985); \textbf{AKARI-FIS} Bright Source Catalogue: 0.8$\arcsec$ rms accurate positions and fluxes at 65 $\mu$m, 90 $\mu$m, 140 $\mu$m, and 160 $\mu$m bands for 427,071 sources over the whole sky with 0.55 Jy sensitivity in 90 $\mu$m band (Yamamura et al. 2010); 

\item \emph{\textbf{Sub-mm/mm.}} \textbf{Planck}: 30–857 GHz (350 $\mu$m to 1 cm) 9-band data for the whole sky, study of the CMBR and detection of 33,566 point sources (Planck 2011); \textbf{WMAP}: 22–-90 GHz fluxes of the whole sky, study of CMBR and detection of 471 point sources (Gold et al. 2011); \textbf{SCUBA}: sub-mm instrument on JCMT providing data in 450 $\mu$m and 850 $\mu$m bands; two large catalogs have been released: Fundamental Map Object Catalog (JCMTSF), 5061 objects, and Extended Map Object Catalog (JCMTSE), 6118 objects (Di Francesco 2008); \textbf{Herschel}: data at 55-–672 $\mu$m with three instruments in 13 wavelength bands and spectra; \textbf{ALMA}: has started providing data and in the nearest future it will produce valuable catalogues in sub-mm/mm wavelengths;

\item \textit{\emph{\textbf{Radio.}}} \textbf{Green Bank 6 cm (GB6)}: positions and 6 cm (4.83 GHz) fluxes for 75,162 sources sensitive to 18 mJy (Gregory et al. 1996); \textbf{NRAO VLA Sky Survey (NVSS)}: positions and 21 cm (1.4 GHz) fluxes for 1,773,484 sources over 33,827 deg$^{2}$, sensitive to 2.5 mJy (Condon et al. 1998); \textbf{Faint Images of the Radio Sky at Twenty centimeters (FIRST)}: 5$\arcsec$ rms accurate radio positions and 21 cm (1.4 GHz) fluxes for 946,432 sources from 10,000 deg2 area of the Northern extragalactic sky, sensitive to 1 mJy (Helfand et al. 2015); \textbf{Sydney University Molonglo Sky Survey (SUMSS)}: 36 cm (843 MHz) fluxes for 211,050 sources over 8000 deg$^{2}$ of the Southern sky, sensitive to 1 mJy (Mauch et al. 2012); \textbf{Westerbork Northern Sky Survey (WENSS)}: 49 cm (610 MHz) and 92 cm (330 MHz) fluxes for 229,420 sources over 9950 deg$^{2}$ of the Northern sky, sensitive to 18 mJy (de Bruyn et al. 1998); \textbf{7 Cambridge (7C)}: 198 cm (151 MHz) fluxes for 43,683 sources over 2400 deg$^{2}$ of the Northern sky, down to 40 mJy (Hales et al. 2007). Deeper fields are observed on small areas (VLA, ATCA, etc.).

\end{itemize}

In Table 1 we give a comparative list of multiwavelength all-sky and large-area surveys.


\begin{table}[!t]
\begin{center}
\vbox{\footnotesize\tabcolsep=3pt
\parbox[c]{124mm}{\baselineskip=10pt
{\smallbf\ \ Table 1.}{\small\
Main data for the most important all-sky and large-area astronomical surveys providing multiwavelength photometric data. Catalogues are given in the order of increasing wavelengths.}}
\begin{tabular}{lclrrrr}
\hline
Survey, & Years & Spectral & Sky area & Sensitivity & Number of & Density \\  
Catalogue &     & range    & (deg$^{2}$)   & (mag/mJy)   & sources   & (obj/deg$^{2}$) \\
\hline
Fermi-GLAST & 2008-2014 & 10MeV-100GeV & All-sky   &            &         3,033 &     0.07 \\
CGRO      & 1991-1999 & 20keV-30GeV    & All-sky   &            &         1,300 &     0.03 \\
INTEGRAL  & 2002-2014 & 15keV-10MeV    & All-sky   &            &         1,126 &     0.03 \\
ROSAT BSC & 1990-1999 & 0.07-2.4 keV   & All-sky   &            &        18,806 &     0.46 \\
ROSAT FSC & 1990-1999 & 0.07-2.4 keV   & All-sky   &            &       105,924 &     2.57 \\
GALEX AIS & 2003-2012 & 1344-2831\AA   & 21,435    & 20.8$^{m}$ &    65,266,291 &  3044.85 \\
APM       & 2000      & opt b, r       & 20,964    & 21.0$^{m}$ &   166,466,987 &  7940.61 \\
MAPS      & 2003      & opt O, E       & 20,964    & 21.0$^{m}$ &    89,234,404 &  4256.55 \\
USNO-A2.0 & 1998      & opt B, R       & All-sky   & 21.0$^{m}$ &   526,280,881 & 12757.40 \\
USNO-B1.0 & 2003      & opt B, R, I    & All-sky   & 22.5$^{m}$ & 1,045,913,669 & 25353.64 \\
GSC 2.3.2 & 2008      & opt j, V, F, N & All-sky   & 22.5$^{m}$ &   945,592,683 & 22921.79 \\
Tycho-2   & 1989-1993 & opt BT, VT     & All-sky   & 16.3$^{m}$ &     2,539,913 &    61.57 \\
SDSS DR12 & 2000-2014 & opt u, g, r, i, z & 14,555 & 22.2$^{m}$ &   932,891,133 & 64094.20 \\
DENIS     & 1996-2001 & 0.8-2.4 $\mu$m & 16,700    & 18.5$^{m}$ &   355,220,325 & 21270.68 \\
2MASS PSC & 1997-2001 & 1.1-2.4 $\mu$m & All-sky   & 17.1$^{m}$ &   470,992,970 & 11417.46 \\
2MASS ESC & 1997-2001 & 1.1-2.4 $\mu$m & All-sky   & 17.1$^{m}$ &     1,647,599 &    39.94 \\
WISE      & 2009-2013 & 3-22 $\mu$m    & All-sky   & 15.6$^{m}$ &   563,921,584 & 13669.83 \\
AKARI IRC & 2006-2008 & 7-26 $\mu$m    & 38,778    & 50 mJy     &       870,973 &    22.46 \\
IRAS PSC  & 1983      & 8-120 $\mu$m   & 39,603    & 400 mJy    &       245,889 &     6.21 \\
IRAS FSC  & 1983      & 8-120 $\mu$m   & 34,090    & 400 mJy    &       173,044 &     5.08 \\
IRAS SSSC & 1983      & 8-120 $\mu$m   & 39,603    & 400 mJy    &        16,740 &     0.42 \\
AKARI FIS & 2006-2008 & 50-180 $\mu$m  & 40,428    & 550 mJy    &       427,071 &    10.56 \\
Planck    & 2009-2011 & 0.35-10 mm     & All-sky   & 183 mJy    &        33,566 &     0.81 \\
WMAP      & 2001-2011 & 3-14 mm        & All-sky   & 500 mJy    &           471 &     0.01 \\
GB6       & 1986-1987 & 6 cm           & 20,320    & 18 mJy     &        75,162 &     3.70 \\
NVSS      & 1998      & 21 cm          & 33,827    & 2.5 mJy    &     1,773,484 &    52.43 \\
FIRST     & 1999-2015 & 21 cm          & 10,000    & 1 mJy      &       946,432 &    94.64 \\
SUMSS     & 2003-2012 & 36 cm          &  8,000    & 1 mJy      &       211,050 &    26.38 \\
WENSS     & 1998      & 49/92 cm       &  9,950    & 18 mJy     &       229,420 &    23.06 \\
7C        & 2007      & 198 cm         &  2,388    & 40 mJy     &        43,683 &    18.29 \\
\hline
\end{tabular}
}
\end{center}
\vskip-8mm
\end{table}

\sectionb{3}{OPTICAL IMAGES, PROPER MOTIONS, AND VARIABILITY}

Beside the main MW catalogues giving photometric data, there have been a number of astronomical surveys aimed at covering large areas and obtaining optical images (atlases), measuring proper motions or variability. Most important among these surveys and catalogues are:

\begin{itemize}

\item \emph{\textbf{Optical Images.}} \textbf{Digitized Sky Survey (DSS) I}: all-sky digitized images in two bands, blue and red from POSS-I (Palomar Observatory Sky Survey) / SERC-J, 1.67$\arcsec$ sampling (McGlynn et al. 1994); \textbf{DSS II}: all-sky digitized images in three bands, blue, red, and IR from POSS-II/AAO-SES, 1$\arcsec$ sampling (Lasker et al. 1996); as mentioned above, most objects are catalogued in USNO B1.0 and GSC 2.3.2; \textbf{SDSS DR12}: digital images for 14,555 deg$^{2}$ in five bands, u, g, r, i, z, 0.1$\arcsec$ resolution (Alam et al. 2015), this is the most accurate large survey; 

\item \emph{\textbf{Proper Motions.}} \textbf{USNO-B1.0}: 0.4$\arcsec$ rms positions and proper motions for all 1,045,913,669 objects over the whole sky (Monet et al. 2003); \textbf{GSC 2.3.2}: 0.3$\arcsec$ rms positions for 945,592,683 objects, proper motions were planned, however due to calibration problems they are not listed (Lasker et al. 2008); \textbf{Tycho-2}: 85 mas rms positions and proper motions for 2,539,913 objects over the whole sky (Hog et al. 2000); besides, proper motions are accessible based on comparison between POSS I epoch combined measurements from APM/MAPS/USNO-A2.0/USNO-B1.0 and POSS II epoch combined measurements from USNO-B1.0/GSC2.3.2 (Mickaelian \& Sinamyan 2010); \textbf{Gaia} will measure accurate positions and proper motions for 1 billion stars with an accuracy of about 20 µas at 15m, and 200 µas at 20m;

\item \emph{\textbf{Variability.}} Such objects are being obtained by repeated observations and are given in a number of catalogues. \textbf{General Catalogue of Variable Stars (GCVS)}: all catalogued 80,671 variables; variability types, typically 0.1m photometric accuracy (Samus et al. 2011); \textbf{All Sky Automated Survey (ASAS)}: ASAS Catalogues of Variable Stars (ACVS) containing some 50,000 stars, including some 30,000 new ones (Pojmanski 1998); \textbf{Northern Sky Variability Survey (NSVS)}: ~14,000,000 objects with 8$^{m}$--15.5$^{m}$ for the area with $\delta$ $>$ –38$^{\circ}$ (33,326 deg$^{2}$); median R ($\sim$0.02$^{m}$ accuracy), measurements scatter, amplitudes, periods, and variability types when possible; 0.04m accuracy (Wozniak et al. 2004a); 8678 catalogued classified red variables from NSVS: Mira type, SR, etc. (Wozniak et al. 2004b); \textbf{International Variable Star Index}: the most complete list of stars (altogether 203,438 objects) ever mentioned as variables (www.aavso.org/vsx); besides, variability (as proper motion) studies are possible based on comparison between POSS I epoch combined measurements from APM/MAPS/USNO-A2.0/USNO-B1.0 and POSS II epoch combined measurements from USNO-B1.0/GSC2.3.2 (Mickaelian et al. 2011); \textbf{Catalina Sky Survey (CSS)}: Cataclysmic Variables (Drake et al. 2014a), Periodic Variable Stars (Drake et al. 2014b) and RR Lyrae stars (Abbas et al. 2014); \textbf{Panoramic Survey Telescope and Rapid Response System (Pan-STARRS)}: having its major goal to discover and characterize Earth-approaching objects, both asteroids and comets, that might pose a danger to our planet, Pan-STARRS discovers many variable stars, as it performs repeated observations. 
\end{itemize}

Summarizing, we have many times covered the entire sky with optical imaging thus providing data for proper motion and variability measurements. More than 200,000 variable objects have been discovered.

\sectionb{4}{SPECTROSCOPIC SURVEYS AND CATALOGUES}

Spectroscopic surveys provide important data that can be used for detailed studies of objects. SDSS gives both photometric and spectral data in large area. Most important among these surveys and catalogues are:

\begin{itemize}

\item \emph{\textbf{Objective Prism Spectroscopy.}} Large amount of spectra have come from objective prism surveys. \textbf{First Byurakan Survey (FBS or Markarian Survey)}: 17,000 deg$^{2}$ in the Northern sky, 20,000,000 low-dispersion spectra (Markarian et al. 1989); \textbf{Second Byurakan Survey (SBS)}: 965 deg$^{2}$ in the Northern sky, some 3,000,000 low-dispersion spectra (Stepanian 2005); \textbf{Case Low-Dispersion Northern Sky Survey }(Pesch et al. 1995); \textbf{Hamburg Quasar Survey (HQS)}: 14,000 deg$^{2}$ in the Northern sky (Hagen et al. 1999); \textbf{Hamburg-ESO Survey (HES)}: 9,000 deg$^{2}$ in the Southern sky (Wisotzki et al. 2000); digitized copies of FBS, HQS, and HES are available online (Mickaelian et al. 2007; Hagen et al. 1999; Wisotzki et al. 2000, respectively); thus, objective prism images for most of the extragalactic sky, 17,000 deg$^{2}$ in North and 9,000 deg$^{2}$ in South are available; 

\item \emph{\textbf{Medium Dispersion Spectroscopy.}} \textbf{2-degree Field and 6-degree Field (2dF/6dF)}: medium dispersion 3700-7900\AA ~spectra for 346,061 galaxies and 49,425 stellar objects, including 23,660 QSOs (galaxy redshift and QSO surveys) (Colless et al. 2001; Croom et al. 2004); 

\item \emph{\textbf{Digital Spectroscopy.}} \textbf{SDSS}: 3800-–9200\AA\AA ~R=1800-–2200 spectra for 4,355,200 selected objects (mainly galaxy redshift and QSO surveys), including 2,401,952 galaxies, 477,161 QSOs, and 851,968 stars (Alam et al. 2015); \textbf{Calar Alto Legacy Integral Field Area (CALIFA)} Survey: is mapping 600 galaxies with imaging spectroscopy (IFS) and produces more than 1 million spectra; \textbf{Galaxy and Mass Assembly (GAMA)} spectroscopic survey: is for 300,000 galaxies and also provides millions of spectra. 

\end{itemize}

At present some 5 million objects (compared to 300,000 ones 20 years ago) have spectroscopy giving understanding on their nature and possibility for detailed investigation. The number of QSOs doubles every 5 years (V\'eron-Cetty \& V\'eron 2010; P\^aris et al. 2014). Thousands of Blazars have been identified (Massaro et al. 2015).

A comparative table (Table 2) gives an understanding on various parameters of low-dispersion objective prism surveys and SDSS.


\begin{table}[!t]
\begin{center}
\vbox{\footnotesize\tabcolsep=3pt
\parbox[c]{124mm}{\baselineskip=10pt
{\smallbf\ \ Table 2.}{\small\
Main characteristics of major low-dispersion surveys and SDSS. Most of them are extragalactic surveys so that mainly high galactic latitudes are covered both in the North and South.}}
\begin{tabular}{lllcrcccl}
\hline
Survey & Years & Equipment & Emuls. & D at & Spectral  & Covered & V$_{lim}$ & Objects of \\
       &       &          &           & H$\gamma$ & Range, \AA & Area    &           & Interest \\
\hline
FBS  & 1965- & 102/132cm & IIa-F & 1800 & 3400-6900 & $|b|$$>$15$^{\circ}$ & 17.5 & UVX (Mkn) \\
     & 1980  & Schmidt,      & & & & $\delta$$>$-15$^{\circ}$ & & galaxies, BSOs, \\ 
     &           & 1.5$^{\circ}$ prism & & & & 17,000 deg$^{2}$ & & 20 mln spectra \\
SBS  & 1978- & 102/132cm & IIIa-J & 1800 & 3400-5300 & $|b|$$>$30$^{\circ}$ & 19 & UVX galaxies, \\
     & 1991  & Schmidt,      & IIIa-F &  900 & 4950-5400 & 49$^{\circ}$$<$$\delta$$<$61$^{\circ}$ & & QSO/Sy, BCDG, \\ 
     &           & 1.5$^{\circ}$/3$^{\circ}$/4$^{\circ}$ & IV-N & 280 & 6300-6950 & 965 deg$^{2}$ & & hot stars \\
Case & 1983- & 61/91cm & IIIa-J & 1350 & 3400-5300 & $|b|$$>$30$^{\circ}$ & 18 & BSOs, UVX \\
     & 1995  & Schmidt, & & & & $\delta$$>$30$^{\circ}$ & & galaxies \\ 
     &           & 1.8$^{\circ}$ prism & & & & & & CSO/CBS/CG \\
HQS  & 1985- & 80cm & IIIa-J & 1390 & 3400-5300 & $|b|$$>$20$^{\circ}$ & 19 & QSOs, \\
     & 1997  & Schmidt,        & & & & $\delta$$>$0$^{\circ}$ & & HQS/RASS \\ 
     &           & 1.7$^{\circ}$ prism & & & & 14,000 deg$^{2}$ & & \\
HES  & 1990- & 1m Schmidt, & IIIa-J & 280 & 3400-5300 & $|b|$$>$30$^{\circ}$ & 18 & QSOs \\
     & 1996  & 4$^{\circ}$ prism & & & & $\delta$$<$+2.5$^{\circ}$ & & hot stars \\
     &           &                   & & & & 9,000 deg$^{2}$ & & \\
SDSS & 2000- & 2.5m R-C & CCD & res.: & 3800-9200 & $|b|$$>$30$^{\circ}$ & 21 & 1 bln objects; \\
     & 2020  & Double MOS &     & 2.5\AA & & $\delta$$>$0$^{\circ}$ & & 4 mln spectra, \\
     &           &           &     &       & & 14,555 deg$^{2}$ & & 450,000 QSOs \\
\hline
\end{tabular}
}
\end{center}
\vskip-8mm
\end{table}

\sectionb{5}{BYURAKAN SURVEYS AND THEIR DIGITIZED VERSIONS}

The \textbf{First Byurakan Survey (FBS)} has been carried out by Markarian, Lipovetski and Stepanian in 1965-1980 with BAO 102/132/213 cm (40$\arcsec$/52$\arcsec$/84$\arcsec$) Schmidt telescope with 1.5$^{\circ}$ prism (Markarian et al. 1989). 2050 Kodak IIAF, IIaF, IIF, and 103aF photographic plates in 1133 fields (4$^{\circ}$$\times$4$^{\circ}$ each, the size being 16 cm $\times$ 16 cm) have been taken. FBS covers 17,000 deg$^{2}$ of all the Northern sky and part of the Southern sky ($\delta$ $>$ -15$^{\circ}$) at high galactic latitudes ($|$b$|$$>$15$^{\circ}$). In some regions, it even goes down to $\delta$ = -19$^{\circ}$ and $|$b$|$=10$^{\circ}$. The limiting magnitude on different plates changes in the range of 16.5$^{m}$--19.5$^{m}$ in V, however for the majority it is 17.5$^{m}$--18$^{m}$. The scale is 96.8$\arcsec$/mm and the dispersion is 1800 \AA/mm near H$\gamma$ and 2500 \AA/mm near H$\beta$ (mean spectral resolution being about 50\AA). Low-dispersion spectra cover the range 3400\AA-6900\AA, and there is a sensitivity gap near 5300\AA, dividing the spectra into red and blue parts. It is possible to compare the red and blue parts of the spectrum (easily separating red and blue objects), follow the spectral energy distribution (SED), notice some emission and absorption lines (such as broad Balmer lines, molecular bands, He, N$_{1}$+N$_{2}$ lines, broad emission lines of QSOs and Seyferts, etc.), thus making up some understanding about the nature of the objects. The FBS is made up of zones (strips), each covering 4$^{\circ}$ in declination and all right ascensions except the Galactic plane regions. In all there are 28 zones, which are named by their central declination (ex. zone +27$^{\circ}$ covers +25$^{\circ}$$<$$\delta$$<$+29$^{\circ}$, zone +63$^{\circ}$ has +61$^{\circ}$$<$$\delta$$<$+65$^{\circ}$, etc). The zones and the neighboring plates in right ascension overlap about 0.1$^{\circ}$, as the exact size of a plate is 4.1$^{\circ}$$\times$4.1$^{\circ}$, thus making the whole area complete. Each FBS plate contains low-dispersion spectra of some 15,000--20,000 objects, and there are some 20,000,000 objects in the whole survey.
 
In 1988, Mickaelian continued the survey to low galactic latitudes to check if it is possible to work with the low-dispersion spectra in these crowded regions. Two regions of the Milky Way of the zones +39$^{\circ}$ and +43$^{\circ}$ have been covered. 28 Kodak IIaF, IIIaF and 103aF plates in 11 and 8 fields having 171 deg$^{2}$ and 117 deg$^{2}$ surface respectively have been obtained. These plates are especially useful for discovery of white dwarfs and other galactic objects.

Main features of FBS are:

\begin{itemize}

\item First systematic objective-prism survey in the history of astronomy,
\item The largest objective-prism survey of the Northern sky (17,000 deg$^{2}$),
\item New method of search for active galaxies,
\item Revelation of 1515 UVX galaxies: some 300 AGN and some 1000 HII galaxies,
\item Classification of Seyferts into Sy1 and Sy2 types,
\item Definition of Starburst (SB) galaxies,
\item FBS Blue Stellar Objects (BSOs) and Late-type Stars,
\item Optical identification of IRAS galaxies (BIG and BIS objects); discovery of new AGN and ULIRGs.

\end{itemize}

The \textbf{Second Byurakan Survey (SBS)} was carried out by Markarian, Stepanian, Erastova, Lipovetski, Chavushian, and Balayan in 1978-1991 with BAO 102/132/213cm Schmidt telescope with 1.5$^{\circ}$, 3$^{\circ}$ and 4$^{\circ}$ prisms, in combination with hypersensitized Kodak IIIaJ and IIIaF plates and filters, giving different ranges of spectrum (3500\AA-5400\AA, 4950\AA-5400\AA and 6300\AA-6950\AA, respectively) (Markarian \& Stepanian 1983; Stepanian 2005). In total, 550 plates have been obtained in 65 fields, 4$^{\circ}$$\times$4$^{\circ}$ each. SBS covers 965 deg$^{2}$ area at high galactic latitudes ($|b|>$30$^{\circ}$) with 07$^{h}$43$^{m}$$<$$\alpha$$<$17$^{h}$15$^{m}$ and +49$^{\circ}$$<$$\delta$$<$+61$^{\circ}$. The limiting magnitude differs in the range of 18$^{m}$--20$^{m}$ in V, and the survey is complete to 17.5$^{m}$. 7 fields (covering 112 deg$^{2}$) have limiting magnitudes 19.5$^{m}$--20$^{m}$ (so-called Selected or Deep Fields of the SBS), and are of special interest. The dispersion near H$\gamma$ for spectra obtained with 1.5$^{\circ}$, 3$^{\circ}$ and 4$^{\circ}$ prisms is 1800\AA/mm, 900\AA/mm, and 280\AA/mm, respectively (and 1000\AA/mm near H$\alpha$ for 4$^{\circ}$ prism). SBS covers 3 zones of the FBS, with central declinations +51$^{\circ}$, +55$^{\circ}$ and +59$^{\circ}$. In average, each SBS plate contains low-dispersion spectra of some 50,000 objects, and there are some 3,000,000 objects in the whole survey.


\begin{figure}[!tH]
\vbox{
\centerline{\psfig{figure=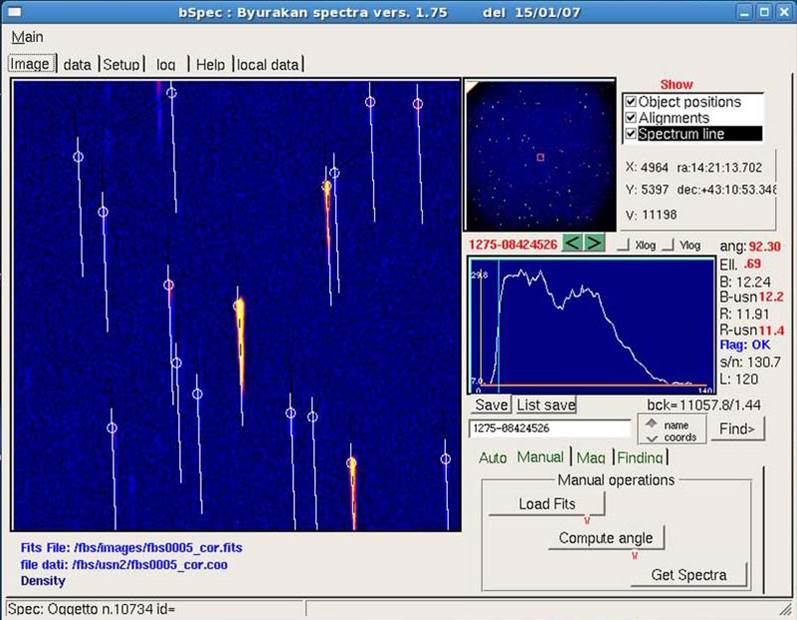,width=70mm,angle=0,clip=}}
\vspace{1mm}
\captionb{2}
{DFBS extraction and analysis software bSpec. It finds objects by an input catalogue data and makes mass extraction and storage of all data with necessary measurements.}
}
\end{figure}


\begin{figure}[!tH]
\vbox{
\centerline{\psfig{figure=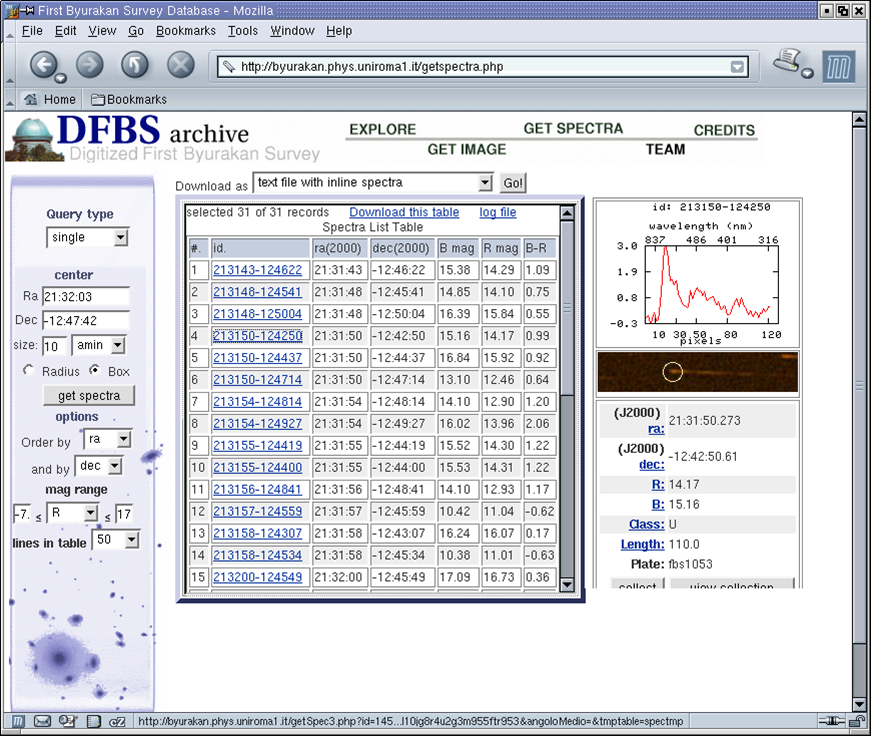,width=70mm,angle=0,clip=}}
\vspace{1mm}
\captionb{3}
{DFBS web interface in mode ``Get Spectra". Spectra may be explored and/or extracted in single or batch modes (by the given list).}
}
\end{figure}

The \textbf{Digitized First Byurakan Survey (DFBS)} is the digitized version of the famous Markarian Survey. It is the largest spectroscopic database in the world, providing low-dispersion spectra for 20,000,000 objects. DFBS is a joint project of the Byurakan Astrophysical Observatory (BAO), Cornell University (USA) and Universita di Roma ``La Sapienza" (Italy). The whole Northern sky and part of the Southern sky at high galactic latitudes have been observed in FBS, altogether more than 17,000 deg$^{2}$. It is especially valuable for extragalactic research. 1500 UV-excess galaxies (Markarian galaxies), 1100 blue stellar objects and 1050 late-type stars have been discovered on the basis of FBS, as well as 1600 infrared (IRAS) sources have been optically identified. The DFBS has been created in 2002-2005 as a result of digitization and reduction of some 2000 FBS plates. High-accuracy (1$\arcsec$ rms) astrometric solution has been made for each plate. Dedicated software allows quick access to any field by given position and extraction of the needed spectra, their calibration, classification and study. The DFBS is free for the astronomical community. It occupies 360 GB space (85 DVDs). The DFBS catalogue and database are being maintained in Armenia and will be also available at CDS, Cornell and Rome. DFBS is the largest Armenian astronomical database and one of the largest in the world. Some projects based on the DFBS have been already put forward, including search for new QSOs and other AGN, continuation of the FBS Second Part, identifications of IR and X-ray sources, etc. The DFBS is the basis for the Armenian Virtual Observatory (ArVO), which will unify all available astronomical data in Armenia, including all Byurakan archive and data from Byurakan telescopes. ArVO is part of the International Virtual Observatories Alliance (IVOA). SBS has also been partially digitized and DSBS is being created.

In Figure 2 we give the DFBS extraction and analysis software bSpec and in Figure 3, DFBS web interface in mode ``Get Spectra", where one can select spectra from the list and retrieve corresponding data individually or in batch mode, including 1D FITS spectra.

\sectionb{6}{ASTRONOMICAL SURVEYS AND BIG DATA}

Astronomical surveys give so much information that huge catalogues, dedicated archives and databases are being built to store, maintain and use these Big Data (Mickaelian 2015). At present astronomers deal with the following numbers in various wavelength ranges (Table 3), and these numbers increase exponentially. It is estimated that there are some 400 billion stars in the Milky Way galaxy and some 100 billion galaxies in the Universe, so that we are very far to catalogue all these objects. Even after Gaia space mission we will have much more accurate astrometric and photometric data for the stars but not much more completeness of detections. 

Figure 4 shows the distribution of the numbers of discovered astronomical objects by wavelength range; it is obvious that optical studies are well in front of astronomy together with NIR/MIR (due to 2MASS and WISE), so that even the logarithmic scale does not show the small bars corresponding to other wavelength ranges, beside UV, optical, and NIR/MIR. However MW astronomy was born in the recent decades and makes huge steps toward the overall understanding of the Universe with its various manifestations from $\gamma$-ray to radio and in the nearest future most of the objects (e.g. in our Galaxy or all galaxies in the Local Universe) will have their counterparts in all wavelengths. 


\begin{table}[!t]
\begin{center}
\vbox{\footnotesize\tabcolsep=3pt
\parbox[c]{124mm}{\baselineskip=10pt
{\smallbf\ \ Table 3.}{\small\
Number of catalogued sources at different wavelength ranges giving a comparative understanding about the wavelength coverage of the observed Universe.}}
\begin{tabular}{llr}
\hline
Wavelength range & Major surveys/catalogues & Number of          \\
                 &                          & catalogued sources \\
\hline
$\gamma$-ray & Fermi-GLAST, INTEGRAL, Swift        &        10,000 \\
X-ray        & ROSAT, XMM-Newton, Chandra          &     1,500,000 \\
UV           & GALEX, HST                          &   100,000,000 \\
Optical      & SDSS, DSS I, DSS II                 & 1,700,000,000 \\
NIR          & 2MASS, DENIS                        &   600,000,000 \\
MIR          & WISE, AKARI-IRC, SST                &   600,000,000 \\
FIR          & IRAS, AKARI-FIS, SST, Herschel      &       500,000 \\
sub-mm/mm    & Planck, WMAP, SCUBA, Herschel, ALMA &       100,000 \\
Radio        & NVSS, FIRST, SUMSS                  &     2,000,000 \\
\hline
\end{tabular}
}
\end{center}
\vskip-8mm
\end{table}


\begin{figure}[!tH]
\vbox{
\centerline{\psfig{figure=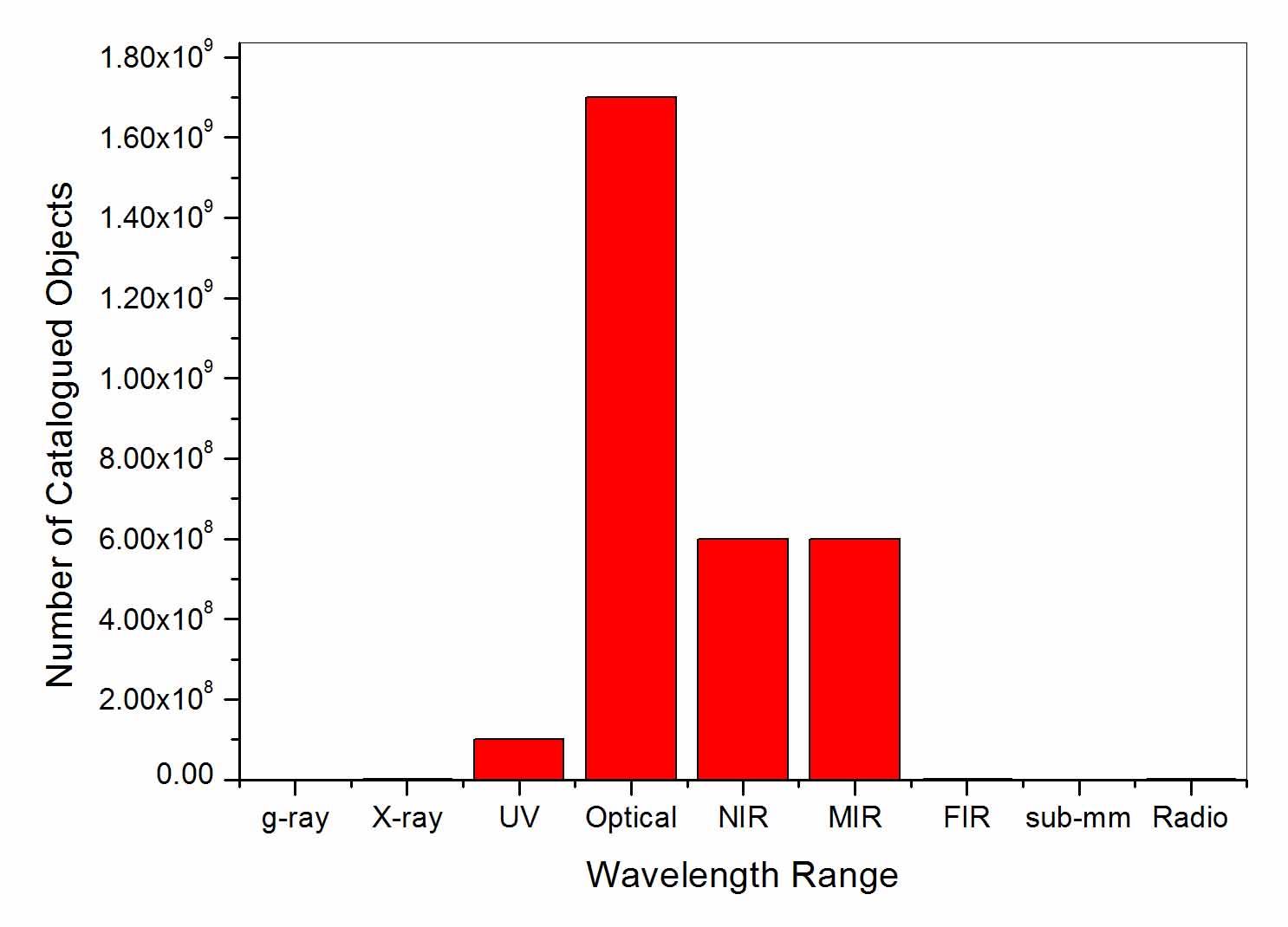,width=80mm,angle=0,clip=}}
\vspace{1mm}
\captionb{4}
{Distribution of the numbers of discovered astronomical objects by wavelength range. Numbers in optical, NIR/MIR and UV are so big that even the logarithmic scale does not show the small bars corresponding to $\gamma$-ray, X-ray, FIR, sub-mm/mm and radio.}
}
\end{figure}

However, establishing correspondence between sources revealed in different wavelengths is a tricky task. Accurate cross-correlations between various MW catalogues are needed to establish genuine counterparts for each object/sourse. Quick cross-matching is being done for almost all catalogues, however; many objects/sources appear to have false associations, as in crowded regions large contamination with other nighboring objects is happening. Very often individual approach should be applied to such associations. Still, a number of cross-correlation software is in use and is being improved.

We give in Figure 5 a comparison of all-sky and large-area astronomical surveys by their wavelength and sky coverage. It is seen that astronomers have covered the whole sky almost in all wavelength ranges. However, Figure 6 shows that not all wavelength have enough deepness, giving a comparison of all-sky/large-area and deep surveys by their sensitivity (limiting magnitude) and number of catalogued objects, which are rather different (only a few other than optical surveys are seen here; 2MASS, DENIS, WISE, GALEX, and IRAS, others having much smaller numbers).

Large astronomical surveys have become one of the most important directions of investigations in our science and they provide the main bulk of information that has been transformed into Big Data and approached astronomy and computer science posing new problems and inquiring new solutions. 


\begin{figure}[!tH]
\vbox{
\centerline{\psfig{figure=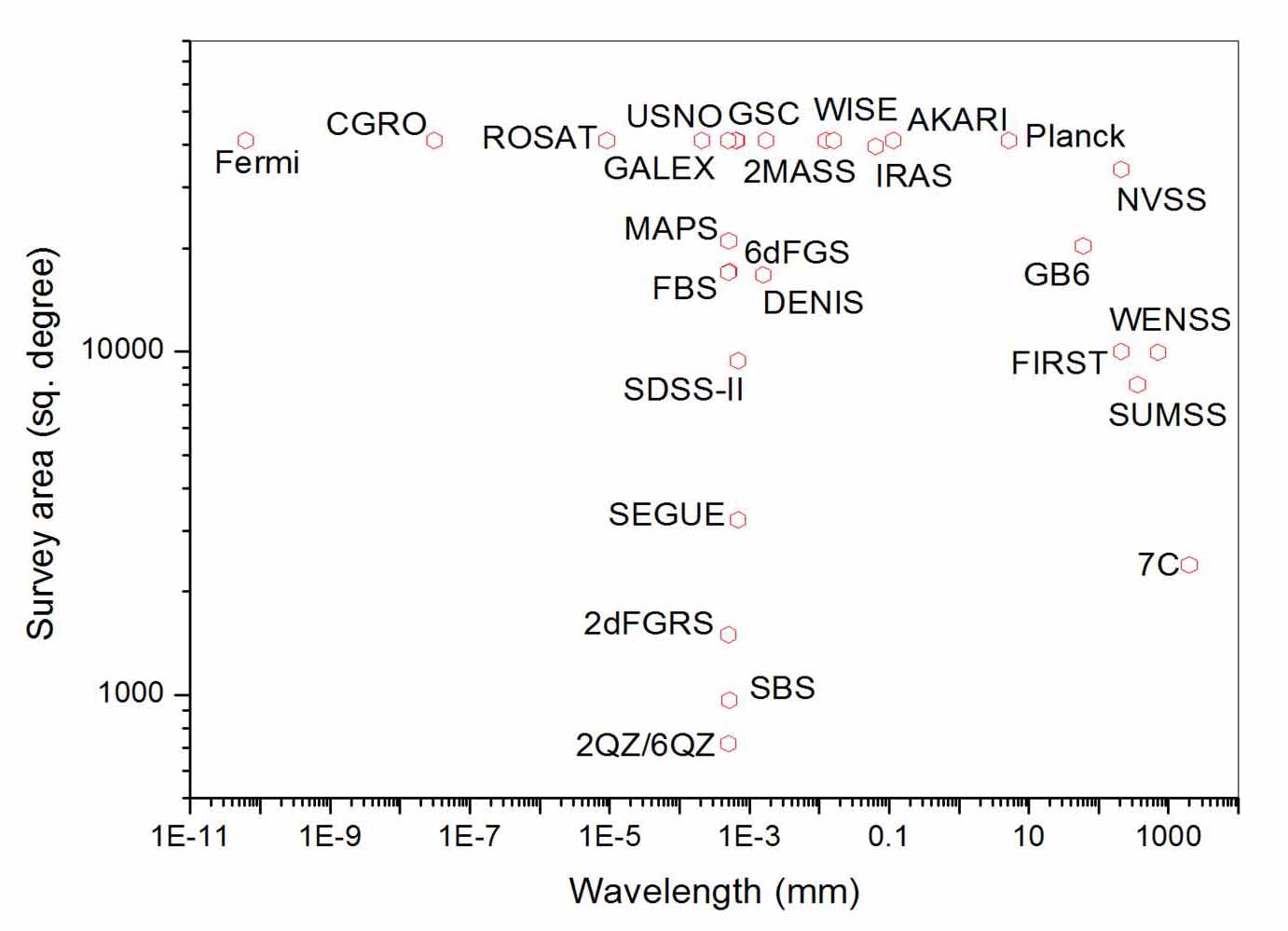,width=80mm,angle=0,clip=}}
\vspace{1mm}
\captionb{5}
{Comparison of all-sky and large-area surveys by wavelength range and survey area. There are all-sky surveys almost in all wavelength domains.}
}
\end{figure}


\begin{figure}[!tH]
\vbox{
\centerline{\psfig{figure=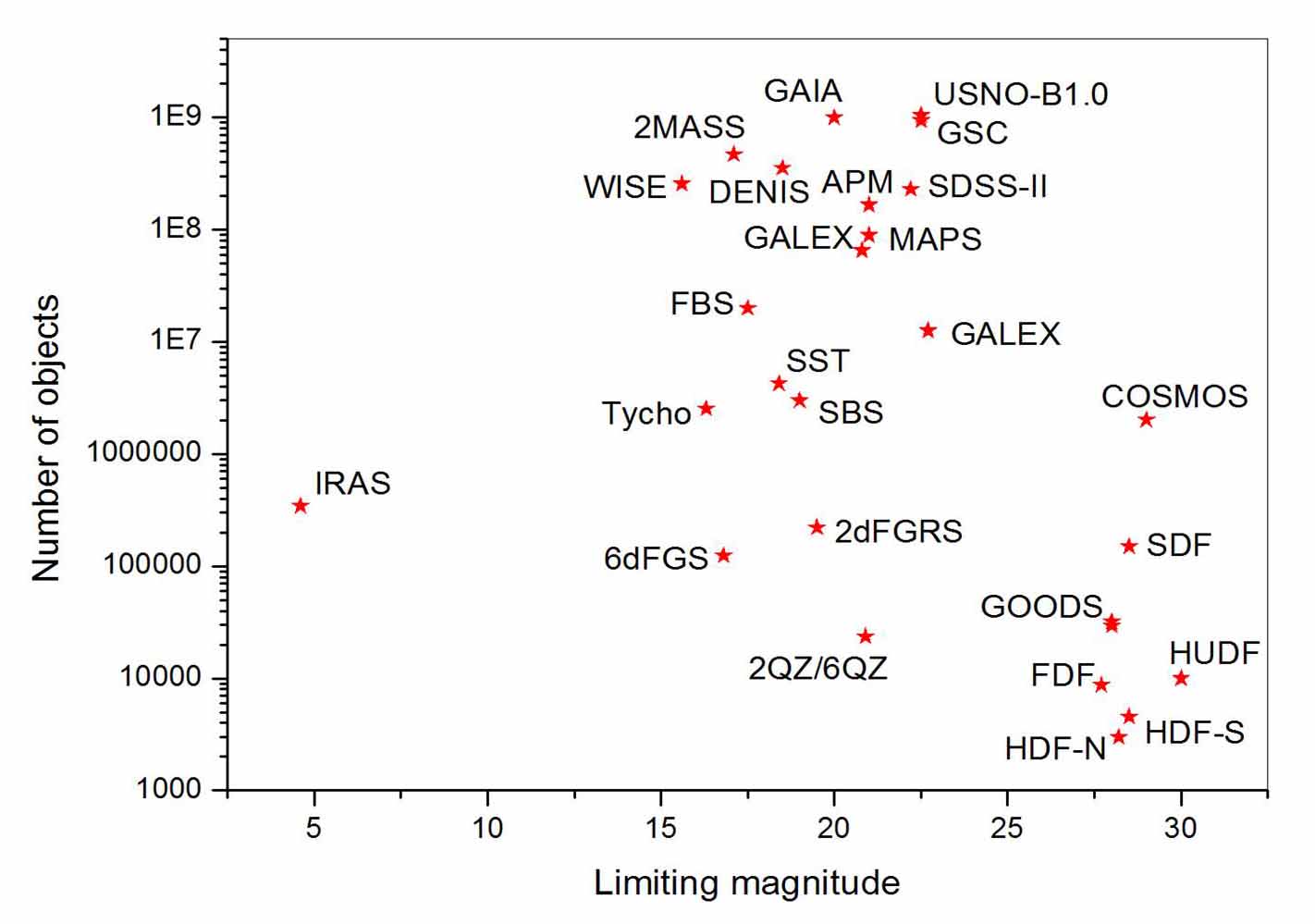,width=80mm,angle=0,clip=}}
\vspace{1mm}
\captionb{6}
{Comparison of all-sky and large-area surveys by limiting magnitude and number of objects. Only a few other than optical surveys are among those having the biggest numbers (hundred millions and billion objects).}
}
\end{figure}

\References

\refb Abbas M. A., Grebel E. K., Martin N. F., et al. 2014, MNRAS 441, 1230

\refb Abrahamyan, H. V.; Mickaelian, A. M.; Knyazyan, A. V. 2015, A\&C 10, 99

\refb Acero, F.; Ackermann, M.; Ajello, M.; et al. 2015, ApJS 218, 23

\refb Alam, S.; Albareti, F. D.; Allende Prieto, C.; et al. 2015, ApJS 219, 12

\refb Bianchi, L.; Herald, J.; Efremova, B.; et al. 2011, Ap\&SS 335, 161

\refb Bird, A. J.; Bazzano, A.; Bassani, L.; et al. 2010, ApJS 186, 1

\refb Cabanela, J. E., Humphreys, R. M., Aldering, G., et al.: 2003 PASP 115, 837

\refb Colless, M.; Dalton, G.; Maddox, S.; 2001, MNRAS 328, 1039

\refb Condon, J. J., Cotton, W. D., Greisen, E. W.; et al.: 1998 AJ 115, 1693

\refb Croom, S. M.; Smith, R. J.; Boyle, B. J.; et al. 2004, MNRAS 349, 1397

\refb Cutri, R. M.; Skrutskie, M. F.; van Dyk, S.; et al. 2003, IPAC/California Institute of Technology

\refb Cutri, R. M.; Wright E. L.; Conrow T.; et al. 2012, WISE All-Sky DR, IPAC/Caltech, VizieR Catalog II/311 

\refb DENIS consortium 2005, 3rd release, Vizier online data catalogue II/263

\refb de Bruyn G., Miley G., Rengelink R., et al. 1998, WENSS Collab. NFRA/ASTRON and Leiden Obs., Vizier online catalogue VIII/62

\refb Di Francesco, J. 2008, AAS 212, 9603

\refb Drake A. J., Gansicke B. T., Djorgovski S. G., et al. 2014a, MNRAS 441, 1186

\refb Drake A. J., Graham M. J., Djorgovski S. G., et al. 2014b, ApJS 213, 9

\refb Gold, B., Odegard, N., Weiland, J.L., et al. 2011, ApJS 192, 15 

\refb Gregory, P. C., Scott, W. K., Douglas, K., Condon, J. J. 1996, ApJS 103, 427 

\refb Hagen, H.-J.; Engels, D.; Reimers, D. 1999, A\&AS 134, 483

\refb Hales, S. E. G.; Riley, J. M.; Waldram, E. M.; et al. 2007, MNRAS 382, 1639

\refb Hartman, R. C.; Bertsch, D. L.; Bloom, S. D.; et al. 1999, ApJS 123, 79

\refb Helfand, D. J.; White, R. L.; Becker, R. H. 2015, ApJ 801, 26

\refb Helou, G.; Walker, D. W. 1985, IRAS small scale structure catalog, JPL, Pasadena, NASA

\refb Hog, E., Fabricius, C., Makarov, V. V., et al. 2000, A\&A, 355, L27

\refb IRAS, 1988, Joint IRAS Scienve Working Group. IRAS PSC, Version 2.0, NASA RP-1190       

\refb Ishihara, D.; Onaka, T.; Kataza, H.; et al. 2010, A\&A 514, 1

\refb Lasker, B. M., Doggett, J., McLean, B., et al. 1996, ASP Conf. Ser. 101, 88       

\refb Lasker, B. M., Lattanzi, M. G., McLean, B. J., et al. 2008, AJ, 136, 735L 

\refb Markarian, B. E.; Lipovetsky, V. A.; Stepanian, J. A.; et al. 1989, Comm. SAO 62, 5

\refb Massaro, E.; Maselli, A.; Leto, C.; et al. 2015, Ap\&SS 357, 75

\refb Mauch T., Murphy T., Buttery H.J., et al. 2012, MNRAS 342, 1117, 2003; Online version 2.1r, 16.02.2012

\refb McGlynn, T., White, N. E., Scollick, K., 1994, ASP Conf. Ser., 61, 34

\refb McMahon, R.G., Irwin, M.J., Maddox, S.J. 2000, The APM-North Catalogue, Inst. of Astron., Cambridge, UK

\refb Mickaelian, A. M. 2012, Baltic Astronomy 21, 331

\refb Mickaelian, A. M. 2015, Astronomy Report, in press

\refb Mickaelian, A. M.; Nesci, R.; Rossi, C.; et al. 2007, A\&A 464, 1177

\refb Mickaelian A.M., Mikayelyan G.A., Sinamyan P.K. 2011, MNRAS 415, 1061

\refb Mickaelian A., Sinamyan P. 2010, MNRAS 407, 681

\refb Monet, D.; Bird, A.; Canzian, B.; et al. 1998, USNO Flagstaff Station and Universities Space Research Association (USRA), Vizier online catalogue I/252 

\refb Monet, D. G.; Levine, S. E.; Canzian, B.; et al. 2003, AJ 125, 984

\refb Moshir, M., Kopan, G., Conrow, T., et al. 1990, IRAS FSC, Version 2.0, NASA  
\refb P\^aris, I.; Petitjean, P.; Aubourg, \'E.; et al. 2014, A\&A 563, 54

\refb Pesch, P.; Stephenson, C. B.; MacConnell, D. J. 1995, ApJS 98, 41 

\refb Planck 2011, Planck Early Release Compact Source Catalogue Planck Collaboration, ESA, 2011, Vizier catalog VIII/88

\refb Pojmanski G. 1998, Acta Astron. 48, 35

\refb Samus', N. N., Durlevich, O. V., Zharova, A. V., et al. 2011, GCVS database, Inst. Astron. and SAI, Moscow

\refb Skrutskie, M. F.; Cutri, R. M.; Stiening, R.; et al. 2006, AJ 131, 1163

\refb Stepanian, J.A. 2005, RMxAA 41, 155

\refb Tsvetkov, M.K., Stavrev, K.Y., Tsvetkova, K.P., et al. 1994, Proc. IAU S161, Kluwer, Dordrecht, p. 359

\refb V\'eron-Cetty, M.-P.; V\'eron, P. 2010, A\&A 518, 10

\refb Voges, W., Aschenbach, B., Boller, T., et al. 1999, A\&A, 349, 389    
\refb Voges, W., Aschenbach, B., Boller, T., et al. 2000, MPE Garching

\refb Wisotzki, L.; Christlieb, N.; Bade, N.; et al. 2000, A\&A 358, 77

\refb Wozniak, P. R., Vestrand, W. T., Akerlof, C. W., et al. 2004, AJ 127, 2436

\refb Yamamura, I.; Makiuti, S.; Ikeda, N.; et al. 2010, AKARI/FIS All-Sky Survey Point Source Catalogues

\end{document}